\documentstyle[aps, preprint,eqsecnum, graphicx, verbatim]{revtex}
\tightenlines
\begin{document}
\title{Strong-weak coupling duality in anisotropic current interactions}
\author{
Denis Bernard $^\P$\footnote{Member of the CNRS, dbernard@spht.saclay.cea.fr} 
 and  
Andr\'e  LeClair $^\clubsuit$ }
\bigskip
\address{$\P$ Service de Physique Th\'eorique \footnote{Laboratoire
    de la Direction des
sciences de la mati\`ere du Commissariat \`a l'\'energie atomique.},
F-91191, Gif-sur-Yvette, France.}
\address{$^\clubsuit$ Newman Laboratory, Cornell University, Ithaca, NY
14853.}
\address{and LPTHE\footnote{Laboratoire associ\'e No. 280 au CNRS},  
4 Place Jussieu, Paris, France.}
\medskip
\date{February  2001}
\maketitle

\def\betaf{{$\beta$eta~}}

\begin{abstract}

The recently proposed all orders \betaf function for
current interactions in two dimensions is further investigated.
By using a strong-weak coupling duality of the \betaf function,
and some added topology of the space of couplings we are able to
extend the flows to arbitrarily large or small scales.   
Using a non-trivial RG invariant we are able to identify 
sine-Gordon, sinh-Gordon and Kosterlitz-Thouless phases.  
We also find an additional phase with cyclic or roaming 
RG trajectories.

\end{abstract}
\vskip 0.2cm
%
%
%
%
\def\dx{\frac{d^2x}{2\pi}}
\def\oti{{\otimes}}
\def\bra#1{{\langle #1 |  }}
\def\lb{ \left[ }
\def\rb{ \right]  }
\def\tilde{\widetilde}
\def\hat{\widehat}
\def\*{\star}
\def\[{\left[}
\def\]{\right]}
\def\({\left(}          \def\BL{\Bigr(}
\def\){\right)}         \def\BR{\Bigr)}
        \def\BBL{\lb}
        \def\BBR{\rb}
%
%
\def\zb{{\bar{z} }}
\def\zbar{{\bar{z} }}
\def\frac#1#2{{#1 \over #2}}
\def\inv#1{{1 \over #1}}
\def\half{{1 \over 2}}
\def\d{\partial}
\def\der#1{{\partial \over \partial #1}}
\def\dd#1#2{{\partial #1 \over \partial #2}}
\def\vev#1{\langle #1 \rangle}
\def\ket#1{ | #1 \rangle}
\def\rvac{\hbox{$\vert 0\rangle$}}
\def\lvac{\hbox{$\langle 0 \vert $}}
\def\2pi{\hbox{$2\pi i$}}
\def\e#1{{\rm e}^{^{\textstyle #1}}}
\def\grad#1{\,\nabla\!_{{#1}}\,}
\def\dsl{\raise.15ex\hbox{/}\kern-.57em\partial}
\def\Dsl{\,\raise.15ex\hbox{/}\mkern-.13.5mu D}
%
%
\def\th{\theta}         \def\Th{\Theta}
\def\ga{\gamma}         \def\Ga{\Gamma}
\def\be{\beta}
\def\al{\alpha}
\def\ep{\epsilon}
\def\vep{\varepsilon}
\def\la{\lambda}        \def\La{\Lambda}
\def\de{\delta}         \def\De{\Delta}
\def\om{\omega}         \def\Om{\Omega}
\def\sig{\sigma}        \def\Sig{\Sigma}
\def\vphi{\varphi}
%
%
\def\CA{{\cal A}}       \def\CB{{\cal B}}       \def\CC{{\cal C}}
\def\CD{{\cal D}}       \def\CE{{\cal E}}       \def\CF{{\cal F}}
\def\CG{{\cal G}}       \def\CH{{\cal H}}       \def\CI{{\cal J}}
\def\CJ{{\cal J}}       \def\CK{{\cal K}}       \def\CL{{\cal L}}
\def\CM{{\cal M}}       \def\CN{{\cal N}}       \def\CO{{\cal O}}
\def\CP{{\cal P}}       \def\CQ{{\cal Q}}       \def\CR{{\cal R}}
\def\CS{{\cal S}}       \def\CT{{\cal T}}       \def\CU{{\cal U}}
\def\CV{{\cal V}}       \def\CW{{\cal W}}       \def\CX{{\cal X}}
\def\CY{{\cal Y}}       \def\CZ{{\cal Z}}

\def\rvac{\hbox{$\vert 0\rangle$}}
\def\lvac{\hbox{$\langle 0 \vert $}}
\def\comm#1#2{ \BBL\ #1\ ,\ #2 \BBR }
\def\2pi{\hbox{$2\pi i$}}
\def\e#1{{\rm e}^{^{\textstyle #1}}}
\def\grad#1{\,\nabla\!_{{#1}}\,}
\def\dsl{\raise.15ex\hbox{/}\kern-.57em\partial}
\def\Dsl{\,\raise.15ex\hbox{/}\mkern-.13.5mu D}
%
%
%
\font\numbers=cmss12
\font\upright=cmu10 scaled\magstep1
\def\stroke{\vrule height8pt width0.4pt depth-0.1pt}
\def\topfleck{\vrule height8pt width0.5pt depth-5.9pt}
\def\botfleck{\vrule height2pt width0.5pt depth0.1pt}
\def\Zmath{\vcenter{\hbox{\numbers\rlap{\rlap{Z}\kern
0.8pt\topfleck}\kern 2.2pt
                   \rlap Z\kern 6pt\botfleck\kern 1pt}}}
\def\Qmath{\vcenter{\hbox{\upright\rlap{\rlap{Q}\kern
                   3.8pt\stroke}\phantom{Q}}}}
\def\Nmath{\vcenter{\hbox{\upright\rlap{I}\kern 1.7pt N}}}
\def\Cmath{\vcenter{\hbox{\upright\rlap{\rlap{C}\kern
                   3.8pt\stroke}\phantom{C}}}}
\def\Rmath{\vcenter{\hbox{\upright\rlap{I}\kern 1.7pt R}}}
\def\Z{\ifmmode\Zmath\else$\Zmath$\fi}
\def\Q{\ifmmode\Qmath\else$\Qmath$\fi}
\def\N{\ifmmode\Nmath\else$\Nmath$\fi}
\def\C{\ifmmode\Cmath\else$\Cmath$\fi}
\def\R{\ifmmode\Rmath\else$\Rmath$\fi}








\def\beq{\begin{equation}}
\def\eeq{\end{equation}}
\def\cg{{\cal G}}
\def\ch{{\cal H}}
\def\barray{\begin{eqnarray}}
\def\earray{\end{eqnarray}}

\section{Introduction}

Current-current interactions in two dimensions arise in 
many important physical systems, for example in  non-Fermi liquids,
Kosterlitz-Thouless transitions, and disordered Dirac fermions.  
For the generally non-integrable anisotropic cases, 
the renormalization group (RG) to low orders in perturbation theory has   
been one of the main tools for understanding these theories.  
Recently, an all-orders \betaf function was proposed\cite{GLM}.  
Some difficulties were encountered in understanding completely
the consequences of these \betaf functions, due mainly to the 
existence of poles\cite{Lec}. In this paper we resolve some of these
difficulties by using the duality pointed out in \cite{Lec} combined
with some added topology in the space of couplings.   We focus only
on the anisotropic $su(2)$ case.        

The $\beta$eta functions were originally  computed 
by using the current-algebra Ward identities to isolate the
$\log$ divergences at each order and summing the perturbative series. 
These series converge in a neighborhood of the origin which
we identify as  the perturbative domain. 
The \betaf functions turn out to be rational functions of the couplings and
possess pole singularities at the boundary of this perturbative domain.
We extend the \betaf functions 
away from this domain by analytical continuation.
Although simple,  this ansatz is a hypothesis which may hide
some dynamical effects, for example the possibility that 
degrees of freedom become
relevant at the singularities. 
Assuming that the proposed \betaf functions 
capture all perturbative contributions, the analytical continuation
prescription
provides a non-perturbative definition of them.
One of the aims  of this paper is to check the consistency
of the proposed \betaf functions upon analytical continuation   
and to decipher the  consequences.

It is  useful to recall a few properties
of the renormalization group  transformations we expect, 
and to discuss pathologies that may be encountered.
RG transformations,  defined for example in the 
Wilsonian scheme   by successive
integrations of the fast modes, induce a map on the space of effective
actions. In this space, of a priori huge dimension, 
the RG transformations may be extended to arbitrarily large
scales. The space of effective actions contains many 
irrelevant directions, corresponding to what are called 
stable manifolds, which are washed out in the continuum limit.
The remaining directions, hopefully finite in number, 
label the continuous field theories if they exist. 
For renormalizable theories the RG transformations 
may be projected onto this parameter space.
The renormalized coupling constants may then be viewed
as the values of these parameters at the renormalization length scale.
For renormalizable theories, this then defines RG flows 
which may be extended to arbitrarily large or small scales
as they simply correspond to space dilatations in the
continuous field theories.

In perturbation theory  one computes \betaf functions
in some domains which may not cover all of  phase space. The sizes of these 
domains, as well as the analytical properties of the \betaf
functions, depend on renormalization prescriptions, i.e. on choices
of coordinates. As a consequence, some RG trajectories cannot be 
extended from  perturbative domains to arbitrarily large scales.
For example, one of the perturbative coordinates may blow up after a
finite scale transformation. This signals that either some relevant
degrees of freedom or some non-perturbative effects
have  been missed, or that the perturbative domains 
have to be extended. Thus, as a rule we shall look for
extensions of the perturbative phase spaces such that all 
RG trajectories can  be followed up to infinitely large scales.
In this situation  we may call the RG phase space complete in analogy
with General Relativity where one looks for geodesically
complete manifolds. Similarly, if the RG flows defined in this way 
cannot be extended to small scales, but break at some  
ultraviolet cut-off, this may indicate the absence of a
non-perturbatively defined renormalizable continuous theory.

\bigskip 

\section{Duality in the \betaf function}

\def\Jb{\bar{J}}

The models we consider are perturbations of the $su(2)_k$ WZW
models, the latter being  conformal field theories 
with $su(2)$ affine symmetry \cite{KZ}.
Let us  normalize the $su(2)$ level $k$ current algebra as follows
\beq
\label{2.1}
J_3(z) J_3(0) \sim \frac{k}{2z^2}  , ~~~~~~~
J_3(z) J^\pm (0) \sim \pm \inv{z} J^\pm (0)
,~~~~~
J^+ (z) J^-(0) \sim \frac{k}{2z^2}  + \inv{z} J_3(0)
\eeq
and similarly for $\Jb$, and   
 consider the anisotropic left-right current-current perturbation:
\beq
\label{2.2}
S  =  S_{su(2)_k}  +  \int \dx \(  g_1 (J^+ \Jb^- + J^- \Jb^+ )
+ g_2 J_3 \Jb_3  \)
\eeq
The \betaf functions proposed in \cite{GLM} are 
\barray
\label{2.3}
\beta_{g_1} &=&  \frac{ g_1 (g_2 - g_1^2 k/4 )}
                   {( 1-k^2 g_1^2/16 )(1+kg_2/4 ) }
\\
\beta_{g_2} &=&  \frac {  g_1^2 (1-kg_2/4)^2  }
                         { (1-k^2 g_1^2/16 )^2 }
\earray
The $su(2)$ invariant isotropic lines correspond to 
$g_1=g_2$, and also to $g_1 = -g_2$ due to the $su(2)$ 
automorphism $J^\pm \to - J^\pm$ which can be performed
on the left-moving currents only sending $g_1 \to - g_1$. 

By rescaling the currents $J\to \sqrt{k} J$ in the isotropic
case one easily sees that
\beq
\label{scaling}
\beta_g  =  \inv{k} F(C_{adj}/k , kg)
\eeq
where $C_{\rm adj}$ is the quadratic Casimir in the adjoint representation,
and $F$ is a function of the combinations $C_{\rm adj}/k , kg$. 
The $\log$ divergences that contribute to the \betaf function 
found in \cite{GLM}  are linear in $C_{\rm adj}/k$ to any order
in perturbation theory, which explains the simple $k$ dependence
in (\ref{2.3}).  The manner in which $\log$ divergences were
isolated in \cite{GLM} led to no corrections which are higher
powers of $C_{\rm adj}/k$, and thus corrections to higher
powers of $1/k$,   however the argument given does
not constitute a proof that there are no such corrections.  
The work \cite{GLM}  also did not deal with potential 
infra-red divergences.   
In this work we do not address these issues but simply 
assume the above \betaf function is correct and see if it
can be made sense of.

In \cite{Lec} various regimes of behavior of the RG flows
were identified.  The main difficulty encountered in 
interpreting these flows was due to the fact that in some 
regimes after a finite scale transformation one flows to the 
poles and one cannot extend the flows reliably beyond the
poles numerically. Another difficulty encountered  in  \cite{Lec}
is  that for some RG trajectories one of the coupling
constants $g_{1,2}$ blows up after a finite scale transformation.
 In this paper we propose a resolution 
of this difficulty based on a strong-weak coupling duality
of the \betaf function and some added topology of the space
of couplings.

Consider the isotropic \betaf function for $g_1=g_2 =g$ 
(for $k=1$) with a double pole at $g=-4$: 
\beq
\label{2.4} 
\beta_g =  \frac{16 g^2}{ (g+ 4)^2 } 
\eeq
It exhibits a duality\cite{Lec}.   Define a dual coupling 
$g^* = 16/{g}$ and let $\beta^* (g^*)=
(\partial g^*/\partial g)\, \beta(g)$, then
\beq
\label{2.6}
\beta^* (g^*) = -\beta (g \to g^*)  
\eeq
Suppose the flow starts at $g<-4$ toward the pole.  Then 
at some scale $r_0$,  $g=g^*= -4 \equiv g_0$, where
$g_0$ is the self-dual coupling.  The duality of the
\betaf function implies that a coupling $g$ at a scale
$r$ flows to $g^*$ at a scale $r_0/r$.  Thus the ultraviolet
and infrared values of $g$ are simply related by duality.   
This can be seen from the analytical solution:
\beq
\label{2.7}
g-g^* + 8 \log (g/g_0) = 16 \log( r/r_0 )
\eeq

The anisotropic \betaf function also exhibits the duality
(\ref{2.6}) with 
\beq
\label{2.8}
g_1^* = \frac{16}{k^2 g_1} , ~~~~~~
g_2^* = \frac{16}{k^2 g_2}
\eeq
If $g_{1,2}(r)$ is a solution of the RG equations, so is
$g^*_{1,2}(r_0/r)$ for any $r_0$.
To simplify the notation, unless explicitly stated
we rescale $g_{1,2}$ to absorb the $k$ dependence.

The precise shape of the RG trajectories can be determined
by using the RG invariant
\beq
\label{Q}
Q = \frac{ g_1^2 - g_2^2}{ (g_2-4)^2 (g_1^2 - 16) }
\eeq
It is simple to show
\beq
\label{Qb}
\sum_g  \beta_g \d_g Q = 0
\eeq
In consistency with the duality of the \betaf functions, one
finds that $Q$ is self-dual, i.e. $Q(g) = Q(g^*)$. 
Eq.(\ref{Q}) may be inverted to express $g^2_1$ as a function 
of $Q$ and $g_2$. This gives the equations for the RG trajectories.
The $\beta$eta function for $g_2$ may be written as a function
of $g_2$ and $Q$ only:
$$
\beta_{g_2}= 16\frac{(g_2^2-16Q(g_2-4)^2)(1-Q(g_2-4)^2)}{(g_2+4)^2}
$$
The remaining pole at $g_2=-4$ reflects the instability of the 
theory at that value of $g_2$; see its bosonized form (\ref{3.2})
in the following section.
The zeroes of $\beta_{g_2}$ are then found to be either on the
$g_2$-axis with $g_1=0$ or at infinity $g_1=\infty$.
\medskip

We propose to use the duality (\ref{2.8}) to extend the flows through the
poles. Namely, if at a scale $r$ the couplings are $g_{1,2}(r)$,
and the flow reaches the pole at a scale $r_0$, then at the scale
$r_0/r$  the couplings are given by $g_{1,2}^* (r)$. 
This procedure for example  allows us to make sense of all trajectories
in the shaded domains in Figure 1.
The beta functions (\ref{2.3}) are ill-defined at the self-dual points
and as a consequence there are many trajectories going through these points.
To resolve this paradox one has to locally change the
description of the phase space. Namely, points of the phase space
will be described by coordinates $(g_1,g_2,Q)$ with the 
identification (\ref{Q}). This is a redundant description
away from the self-dual points but not in the  neighborhood of
these points where it amounts to choosing for
example  $(g_2,Q)$ as coordinates.
This procedure, which is related to the mathematical construction
known as local blowing up, changes slightly the topology of the phase
space.

It turns out one additional ingredient is needed for a global
interpretation of the phase diagram.  
Consider the region $g_2>4, g_1<4$.  Using $Q$, 
one sees that here one is attracted to $g_2 \to \infty$
with $0<g_1 < 4$. 
Using duality arguments, it is easy to convince oneself that
these trajectories reach $g_2=\infty$ in a finite RG time $\log r$.
Without further information one cannot extend the RG flows to
arbitrarily large scales
on the two dimensional plane with coordinates $g_1,\,g_2$. 
Since these trajectories are not flowing to a self-dual point,
one cannot use duality to extend the flows.  
However at $|g_2| = \infty$, the \betaf function satisfies 
\beq
\label{2.9}
\beta(g_1 , g_2) = \beta (g_1, -g_2) ~~~~~~~~~~
{\rm when} ~ |g_2| \to \infty
\eeq
Thus if a flow goes to $g_2 = \infty$ after a finite scale
transformation, one can consistently continue the flow 
at $g_2 = -\infty$. This is also consistent with the fact 
that $Q(g_2) = Q(-g_2)$ at $g_2 = \infty$. 
Thus we propose to identify the
points $g_2 = \pm \infty$ which endows the coupling constant
space with the topology of a cylinder. 
Alternatively, instead of identifying
the two infinities $g_2=\pm \infty$,
one can view the procedure as gluing multiple patches 
of the coupling constant space together at $g_2 = \pm \infty$,
with (\ref{2.9}) as a consistent gluing condition. 
The $+\infty$ of one patch is identified with $-\infty$
of the next patch. In this alternative construction 
the phase space will also acquire a non-trivial topology.

With the above hypotheses all RG flows can be extended to 
arbitrarily large or small length scales.  All flows either
begin on the $g_2$ axis and terminate at $g_1=\infty$,
or vice versa. Since both $g_1=0$ and $g_1=\infty$ are simple zeroes
of the \betaf functions, it takes an infinite RG time for
the trajectories to reach these loci. Consider the flows
that originate from $g_1 = 0$, $0<g_2 < 4$.  They are attracted
to the self-dual point $(g_1 , g_2) = (4,4)$, and then
by duality end up at $g_1 = \infty$, $4<g_2 < \infty$.  
Similarly, the flows that originate from  $g_1=0$, $g_2 < -4$
end up at $g_1 = \infty$, $-4<g_2 <0$. 
The flows that end up on the $g_2$ axis with $-4<g_2<0$
originate from $g_1=\infty$, $-\infty<g_2<-4$
by duality around the self-dual point $(4,-4)$.
Consider now the flows that originate from 
$g_1 = 0$, $g_2 >4$.  After a finite RG time $\log r$, they
are attracted to $g_2 = \infty$ with $0<g_1 < 4$.  
These flows then continue from $g_2 = -\infty$ of the next
patch to the self-dual
point $(g_1, g_2) = (4, -4)$ and end up at 
$g_1=\infty$, $0<g_2 <4$.  Again, since the flow went through
a self-dual point,  the couplings in the UV and IR are related
by duality.   The resulting phase diagram is shown in
figure 1.  For $g_1 <0$, the phase diagram is the reflection
of the figure about the $g_1$ axis since $\beta_{g_1} = 
- \beta(-g_1)$, $\beta_{g_2} = \beta(-g_1)$.  

Interestingly one phase exhibits roaming or cyclic RG trajectories.
Namely, flows that begin in a patch at $g_2= -\infty$, $4<g_1<\infty$,
flow through the poles at $(g_1, g_2) = (4, -4)$, then through
the poles at $(4,4)$, then off to $g_2=\infty$, $4<g_1 < \infty$
where they start over again at $g_2 = -\infty$ in the next patch. 
It takes a finite RG time to go from $g_2=-\infty$ to $g_2=+\infty$
following these trajectories. With the points $g_2=\infty$ and
$g_2=-\infty$ identified,  these trajectories are cyclic, but
in the alternative construction of the phase space by gluing
multiple patches, these trajectories explore  all patches.
Note that these flows pass by the perturbative domain,
$|g_1|<4$, $|g_2|<4$, in which the perturbative expansions of
the \betaf functions converge. If one assumes that 
field theories with initial couplings in the perturbative domain
exist, one is naturally lead to include these RG trajectories
into the phase space.

In the perturbative domain $0<g_{1,2} <4$ one observes flows toward
the $su(2)$ invariant isotropic line, but this line is unstable
beyond the pole and the above duality arguments indicate 
that the $su(2)$ symmetry does not continue to be restored at
strong coupling. The phases above or below the isotropic line
have then very different large scale asymptotics.
 As we describe in the next section,  if the
flows continued along the isotropic line, this would be inconsistent
with the believed existence of a sine-Gordon phase. 
These issues were recently addressed at one loop  in 
\cite{Konik}.

\section{Sine-Gordon and sinh-Gordon phases}

\def\bh{b}

In this section we interpret the RG flows in a bosonized
description. 
When $k=1$, the current algebra can be bosonized as
\beq
\label{3.1}
J^\pm = \inv{\sqrt{2}} \exp \( {\pm i \sqrt{2} \vphi }\) ,
~~~~~ J_3  = \frac{i}{\sqrt{2}} \d_z \vphi
\eeq
where $\vphi (z) $ is the  $z$-dependent  part of a free massless
scalar field $\phi = \vphi (z) + \bar{\vphi} (\bar{z})$.
Viewing the $g_2$ coupling as a perturbation of the kinetic term
and rescaling the field $\phi$ one obtains the sine-Gordon (sG)
action
\beq
\label{3.2}
S = \inv{4\pi} \int d^2 x \[ \inv{2} (\d\phi)^2
+ g_1 \> \cos (\bh \phi ) \]
\eeq
where $\bh (g_1, g_2)$ is to be determined.\footnote{$\bh$ 
is related the conventional sine-Gordon coupling 
$\beta$  by $\bh = \beta/\sqrt{4\pi}$, where $\beta = \sqrt{4\pi}$ 
corresponds to the free-fermion point and $\sqrt{8\pi}$ is
the Kosterlitz-Thouless point.}

When $g_1 \approx 0$, $\beta_{g_1} \approx 4g_2 g_1/(4+g_2)$. 
Since this is linear in $g_1$, we can identify the dimension 
$\Gamma$ of $\cos \bh \phi$. 
Generally,  if $\beta (g_c) = 0$, then
\beq
\label{dim}
\beta (g) = (2-\Gamma) (g-g_c) + ....
\eeq
where $\Gamma$ is the dimension of the perturbing operator.
Thus  $2-\Gamma = 4g_2 /(4+g_2)$.  Since
$\Gamma = \bh^2$ we find
\beq
\label{3.3}
\frac{\bh^2}{2}  =  \frac{1-g_2/4}{1+ g_2/4}, ~~~~~~~{\rm when~}
g_1 \approx 0
\eeq
With this identification, the \betaf function matches 
the known two-loop result\cite{Lec}. 
Since $g_2$ flows under RG, the above formula cannot be
valid for all $g$ since the exact S-matrix for the sG theory
depends on an RG invariant $\bh$ with $0<\bh^2 < 2$. 
It is natural then that in general $\bh$ is a function of
the RG invariant $Q$.  Matching (\ref{3.3}) at $g_1=0$ one
finds 
\beq
\label{3.5}
\frac{\bh^2}{2} = \inv{1+8\sqrt{Q}}
\eeq
We propose the above formula is valid for all $g$.  A non-trivial
check is at $g_1=\infty$ where 
$\beta_{g_1} \approx 16 g_1/(4+g_2)$.  Again identifying 
$\bh^2$ through the dimension of $\cos \bh \phi$, one finds
\beq
\label{3.6} 
\frac{\bh^2}{2} = \frac{g_2-4}{g_2+4} 
~~~~~~~{\rm when }  ~g_1 \approx \infty
\eeq
One sees that the expression (\ref{3.5}) matches both (\ref{3.3})
and (\ref{3.6}) and thus has the right behavior at 
$g_1 = 0,\infty$. This is of course consistent with duality (\ref{2.8}).

The flows that originate from $g_1=0$, $0<g_2<4$, have
$0<\bh^2 <2$;  we thus identify this as a sG phase.
Thus we see that the very existence of a sine-Gordon phase
relies on the existence of the RG invariant $Q$. 
The flows that terminate at $g_1=0$, $-4<g_2<0$, have
$2<\bh^2 < \infty$.  The operator $\cos \bh \phi$ is thus
irrelevant and this is the massless Kosterlitz-Thouless phase. 

The other flows that originate from the $g_2$ axis correspond
to a negative $\bh^2$.  Letting $\bh = i \bh_{shG}$, 
we interpret this as a sinh-Gordon phase (shG) with the 
action (\ref{3.2}) where $\cos \bh \phi \to \cosh \bh_{shG} \phi$. 
This analytical continuation is compatible with the
interpretation of the singularities at $g_2=-4$ as consequences
of instabilities of the kinetic term in eq.(\ref{3.2}).
The region $4<g_2<\infty$ corresponds to $0<\bh^2_{shG} <2$
whereas $g_2 < -4$ corresponds to $2<\bh^2_{shG} < \infty$, so
the whole range of $\bh^2_{shG}$ is covered.  The flows
all originate from the $g_2$ axis since $\cosh \bh_{shG}$ is
relevant for $\bh_{shG}$.   Repeating the above argument we
find
\beq
\label{3.7}
\frac{\bh^2_{shG}}{2} = \inv{8\sqrt{Q} -1 } 
\eeq

Consider now the roaming or cyclic RG flows of the last section.  
In this region $Q$ is negative, $-\infty <Q<0$.  Assuming
it is valid in this regime, eq. (\ref{3.5}) implies
then that $\bh^2$ has both real and imaginary parts so 
that it is neither sine-Gordon nor sinh-Gordon.  
This is actually reminiscent of Zamolodchikov's staircase
model\cite{Zamostair}.  The staircase model corresponds
to $\bh^2_{shG}/(2+\bh^2_{shG}) = 1/2 \pm i \theta_0/2\pi $,
which implies $\bh^2_{shG}/2 = e^{i\alpha}$ where
$\cos \alpha = (1-(\theta_0/\pi)^2 )/ (1+ (\theta_0/\pi)^2 )$.
 From (\ref{3.7}) one sees that for small $\alpha$, $i\alpha = 8
\sqrt{Q}$, which is compatible with $Q$ being negative.  
Of course, more investigation is required for  a better
characterization of this phase.
 
The above flows do not have non-trivial fixed points in
{\it both} 
the UV and IR that can be compared with known results.  However,
by letting $g_1$ be imaginary,  there are such flows\footnote{This
check was suggested to us by Al. Zamolodchikov\cite{Luk}}.  
The resulting imaginary potential sine-Gordon theory 
is known to possess both an ultra-violet and infra-red fixed
point, both of which are $c=1$ gaussian conformal field theories 
at different radius of compactification\cite{Nien}\cite{FSZ}.   
The anisotropic $su(2)$  \betaf function precisely confirms
this.   To see this, let $g_1 \to i g_1$.  There is still
a line of fixed points at $g_1 = 0$.  At small coupling
$Q \approx (g_1^2 + g_2^2)/256$, thus the RG contours
are circles rather than hyperbolas.  This implies that
flows can originate in the ultra-violet along the positive
$g_2$ axis and end up at a different fixed point on the
negative $g_2 $ axis.  The values of $g_2$ in the
UV and IR  can be determined from the RG invariance of $Q$.
Since $Q = (g_2/(g_2 -4))^2/16$ at $g_1 = 0$, one finds
\beq
\label{imag}
\frac{ g_2^{IR}}{g_2^{IR} - 4} = -    
\frac{ g_2^{UV}}{g_2^{UV} - 4}   
\eeq
Let us express this result in terms of the anomalous
dimension of the perturbation at the fixed points.
As discussed above, using eq. (\ref{dim}) 
this gives $\Gamma = 2(4-g_2)/(4+g_2)$ 
at $g_1 = 0$ fixed points.   Using (\ref{imag}) one finds
\beq
\label{uvir}
\Gamma_{IR} = \frac{ \Gamma_{UV}}{\Gamma_{UV} -1 } 
\eeq
This agrees precisely with the statements made in \cite{FSZ}.  
Note that for $1<\Gamma_{UV}<2$ relevant,  $\Gamma_{IR}$ is
irrelevant as it should be.   Since (\ref{uvir}) turns out
to be exact, this strongly suggests there are no further corrections
to the \betaf function.   

For higher $k$ the above results are generalized to the 
fractional supersymmetric sine-Gordon model\cite{FSSG}. 
This serves as a mild check of the $k$ dependence of the
\betaf function.   In these models, the interaction
$\cos \bh \phi$ is replaced with $\psi_1 \bar{\psi}_1 
\cos \bh_{fssg}  \phi$ where $\psi_1$ are $Z_k$ parafermions
of dimension $(k-1)/k$.   The case $k=2$ is the supersymmetric
sine-Gordon model.  Since $\beta_{g_1} \approx 
g_2 g_2/(1+kg_2/4)$ when $g_1 \approx 0$, repeating the
above arguments we find
\beq
\label{fssg1}
\bh^2_{fssg} = \frac{2(4-kg_2)}{k(4+kg_2)},\quad
{\rm  when} \quad g_1 \approx 0
\eeq
>From the \betaf functions we find that the fractional supersymmetric
sine-Gordon regime corresponds to $0<g_2<4/k$ which corresponds to 
$0<\bh^2_{fssg}/2 < 1/k$, in agreement with \cite{FSSG}.  
This is not a strong check of the $k$ dependence since we have
input the dimension of the parafermions by hand.   

\section{Conclusion}

In summary, by using duality and endowing the coupling constant
space with some additional topological properties, we managed
to interpret the RG flows at strong and weak coupling based
on the \betaf functions proposed in \cite{GLM}.  This supports
the analytic continuation of the \betaf function to strong coupling. 

The cyclic or roaming RG trajectory we found needs further investigation.
We point out that cyclic RG flows are ruled out under the
assumptions of the c-theorem which shows that the central charge
$c$ always decreases\cite{ctheorem}.  
Though the action
written in terms of currents is hermitian, the bosonized form 
indicates that in some domains the kinetic term can be negative. 
Thus, presumably  the assumption
of positivity breaks down in the cyclic domain. 
We also remark that cyclic RG trajectories have previously 
occurred in certain problems in nuclear physics\cite{nuclear}.

We have extended the scheme described in this paper to the
\betaf functions for the network model computed in \cite{Lec}
and this will be described in a separate publication.

\section{Acknowledgments}

We wish  to thank P. Argyres,  M. Bauer, K. Gawedzki, P. Lepage,
M. Neubert, Al. Zamolodchikov   and J.B. Zuber for discussions.  
This work is in part supported by the NSF and by the CNRS.

\begin{figure}[htb]
\hspace{18mm}
\includegraphics[width=10cm]{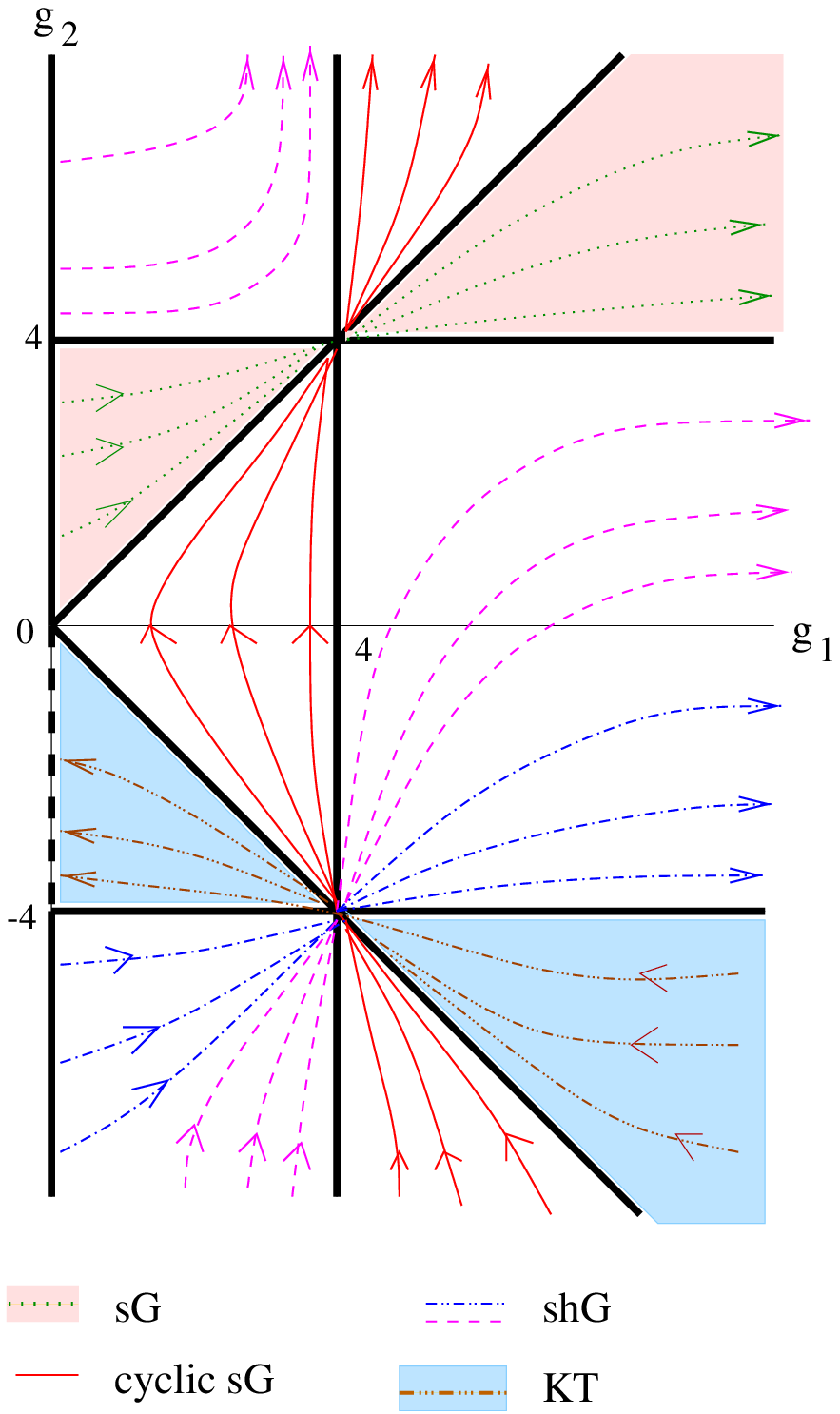}
\caption{Phase Diagram for anisotropic $su(2)$.}
\label{Figure1}
\end{figure}

\end{document}